\documentclass[conference, a4paper]{IEEEtran}
\IEEEoverridecommandlockouts
\usepackage{cite}
\usepackage{amsmath,amssymb,amsfonts}
\usepackage{algorithmic}
\usepackage[utf8]{inputenc}
\usepackage{graphicx}
\usepackage[caption=false]{subfig}
\usepackage{textcomp}
\usepackage{xcolor}
\def\BibTeX{{\rm B\kern-.05em{\sc i\kern-.025em b}\kern-.08em
    T\kern-.1667em\lower.7ex\hbox{E}\kern-.125emX}}

\begin{document}

\newcommand{\vv}{\vspace{2mm}}
\newcommand{\vvv}{\vspace{3mm}}
\newcommand{\vvvv}{\vspace{4mm}}
\newcommand{\hh}{\hspace{2mm}}
\newcommand{\hhh}{\hspace{3mm}}
\newcommand{\hhhh}{\hspace{4mm}}

\title{XAI and Statistical Analysis for Reliable Intrusion Detection in the UAVIDS-2025 Dataset: From Tree to Hybrid and Tabular DNN Ensembles
}

\author{\IEEEauthorblockN{Iakovos-Christos Zarkadis}
\IEEEauthorblockA{
\textit{University of Piraeus}\\
Athens, Greece \\
iakovos.zarkadis@gmail.com}
\and
\IEEEauthorblockN{Christos Douligeris}
\IEEEauthorblockA{\textit{Dept. of Informatics} \\
\textit{University of Piraeus}\\
Piraeus, Greece \\
cdoulig@unipi.gr}
}

\IEEEpubid{979-8-3195-2958-9/26\$31.00 \copyright\ 2026 IEEE}

\maketitle
\begin{abstract}
During the last few years, the term Mechanistic Interpretability, a specific area, under the umbrella of explainable artificial intelligence (XAI), has been introduced, to explain the decisions made by complex machine learning (ML) models in critical systems like UAV intrusion detection systems (UAVIDS). In this paper, we apply best-practices for data pre-processing and examine a wide range of tree-ensembles, deep neural networks, hybrid stacking models and the latest ensemble neural networks to detect intrusions in UAV, with stratified 10-fold cross validation. With our top-performing model, XGBoost, we proceed to Shapley Additive explanations (SHAP), to analyze the global and local feature importances and understand which features, each attack targets, to mimic normal traffic and where the misclassifications occur. Furthermore a distribution analysis follows, by visually comparing violin plots and the curves of kernel density estimations. With the Westfall-Young permutation test for multiple comparisons, the Bandwidth optimization of the KDEs and the selection of Jensen-Shannon Distance for the test, we discover the true causes of false predictions, observed in Wormhole and Blackhole attacks in UAVIDS-2025. The findings provide robust, reliable and explainable models for UAV intrusion detection, along with statistical insights, which capture and clarify the masked nature of the attacks, regarding the challenge of Density Support Intersection, between these attacks, in this dataset.
\end{abstract}

\begin{IEEEkeywords}
Machine learning (ML), explainable artificial intelligence (XAI), unmanned aerial vehicle network intrusion detection systems (UAVIDS), deep neural networks (DNN), Shapley Additive explanation (SHAP).
\end{IEEEkeywords}

\section{Introduction}
Explainable techniques have gained recently everyone's attention, due to their diversity and wide area of application. UAV intrusion detection, has gathered the attention of the research society for it's many applications and dangers, that may occur in IoT networks in combination with the rise of 6G networks and their AI integration \cite{b2}, \cite{b13}, \cite{b19}. XAI methods, such as SHAP, LIME, Grad-CAM have been implemented in research papers, regarding UAV and IoT intrusion detection \cite{b5}, \cite{b8}, \cite{b11}, \cite{b12}, \cite{b14}. Methods like these, have been applied, for a great variety of algorithmic families, such as, Tree-based models, like Decision and Extra trees \cite{b1}, \cite{b20}, ensemble models, like XGBoost, LightGBM \cite{b9}, \cite{b11}, \cite{b12}, neural networks and deep learning, like MLP, RNN, CNN, LSTM \cite{b24}. Due to the critical role of IoT networks, many frameworks have been designed, that combine deep, ensemble and federated learning to enhance intrusion detection \cite{b3}, \cite{b23}. Modern frameworks direct their focus in explainability of predictions, of different modalities and models \cite{b5}, \cite{b14}, \cite{b21}, \cite{b26}, \cite{b28}, \cite{b29}. Other newer frameworks focus on explainable Agentic AI \cite{b27}. 
\par Our focus in this paper, is first, to apply best pre-processing practices to ensure high data-quality, to have reliable, faithful and explainable results that provide deep insight about our data's true nature. We apply a great variety of, well known, state-of-the-art tree-ensembles, like XGBoost \cite{b20}, LightGBM, Histogram-Based Gradient Boosting, Random Forest, hybrid ensembles that combine linear, tree and bayesian models \cite{b10}, deep neural networks, like RealMLP and deep neural network ensembles, like Ensemble-RealMLP, from some of the latest frameworks, such as AutoGluon and PyTabKit, designed specifically for tabular data \cite{b4}, \cite{b15}. 
\IEEEpubidadjcol
\par After we select our best classifier, XGBoost, we proceed to explainable techniques, with SHAP \cite{b22}, like local and global feature importances, in order to analyze how his internal mechanism makes predictions and optimize his decisions. Next we examine the shape of our data, through Box-Plots, per attack, but also their densities and distribution shapes, with violin and  Kernel Density Estimations \cite{b25}, to better understand their univariate nature, along with the challenges that they bring up. Clear evidences are provided, for specific true and false predictions, of selected observations. In the end to really understand and explain the major issue with this dataset, Density Support Intersection, between Blackhole and Wormhole attacks, in most of their feature space, we apply a robust statistical non-parametric Westfall-Young test, for a total of $B=1000$ permutations \cite{b7}, \cite{b16}, \cite{b17}, with bandwidth-optimized KDEs. For this test we calculate Family-Wise Error Rate adjusted $p-values$, to ensure statistical significance, for our multiple comparisons, thus avoiding the Bonferroni correction issues. The statistical function of the test is the Max-Statistic of the Jensen-Shannon distances obtained from the permutations. The chosen procedure provides reliable examination and detection of Support Intersection scenarios, for masked attacks in UAV networks. 
\par This work is organized in two sections. In the first we present the pre-processing steps and the UAV intrusion detection results. In the second we apply SHAP analysis to provide global and local explainability and the non-parametric test, with the KDEs, to provide probabilistic evidence of the major challenge of Density Support Intersection, faced in UAVIDS-2025.

\section{Machine Learning-Based UAVIDS Methodology}

\subsection{Data Collection}
For all the results that will be presented in the next sections we used the UAVIDS-2025 dataset, a recently uploaded tabular dataset for unmanned aerial vehicle network intrusion detection, which contains around $120,000$ observations, $22$ features and the target variable \cite{b12}, \cite{b18}. The target variable has four types of attacks (Sybil, Blackhole, Wormhole, Flooding) and one for the normal traffic. The dataset is not heavily unbalanced, so there is no need for synthetic sampling techniques.

\subsection{Data Pre-Processing}
The pre-processing procedure started by omitting the duplicate values, then removing the redundant feature "Protocol", since the only protocol, which exists is "UDP". In addition, we removed the indexing variable, "FlowID", since it is not a real variable, due to the bias and disruption, which introduces, in Tab. IV of \cite{b12}, as seen by the feature importances. After-that followed, the label-encoding of the target variable. Furthermore, we performed dummy encoding for the source and destination ports and frequency encoding for the destination and source IP address. Next the data were splitted into $80/20$ train/test, in a stratified way, with respect to the target variable, the label. Then we engineered a few features with power and reciprocal transformations. After discovering outliers, with IQR and Box-Plots the Robust scaler was selected. 
\par
\begin{figure}[htbp]
\centerline{\includegraphics[width=0.55\columnwidth]{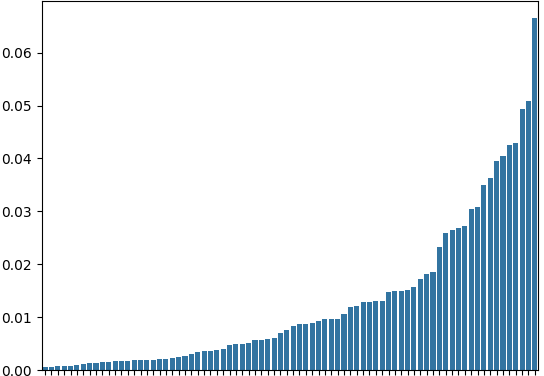}}
\caption{RFE for the Random Forest}
\label{fig1}
\end{figure}

 To conclude to the most important features, we used Recursive Feature Elimination with a Random Forest and obtained the feature importance plot in Fig.~\ref{fig1}. The ones selected from RFE, where those with a percentage of contribution higher than $2.5\%$.

\begin{figure}[htbp]
     \centering
     \subfloat[Pearson\label{fig:pearson}]{%
         \includegraphics[width=0.23\textwidth]{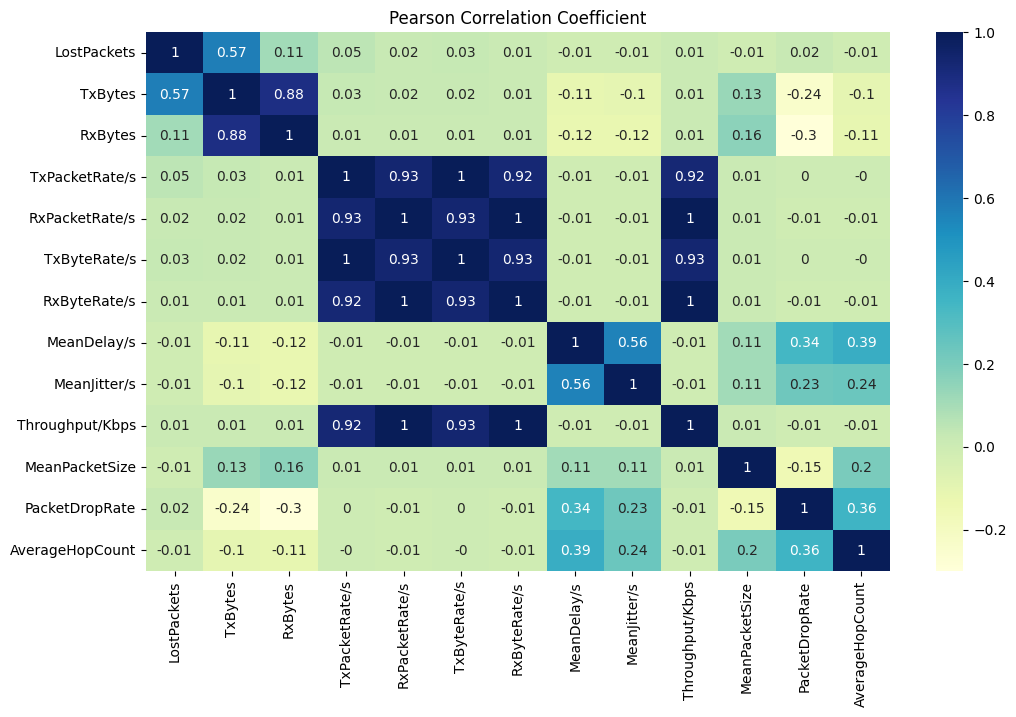}%
     }%
     \hspace{2pt} 
     \subfloat[Kendall\label{fig:kendall}]{%
         \includegraphics[width=0.23\textwidth]{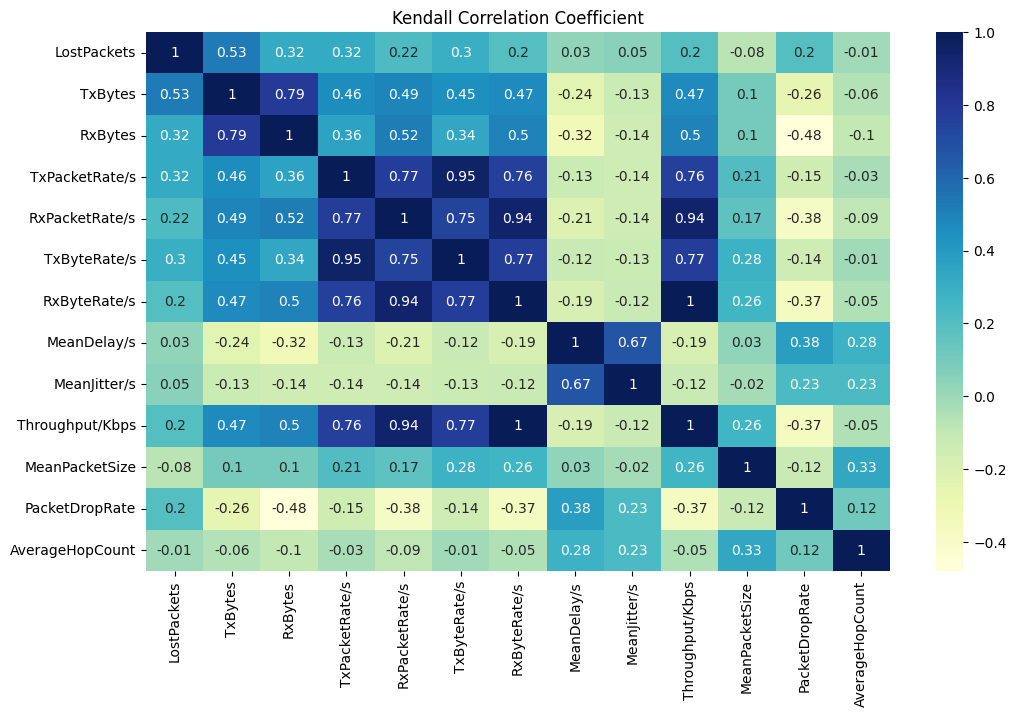}%
     }%
     \caption{Correlation Coefficients after Feature Selection}
     \label{fig:fig2}
\end{figure}

From the feature selection, the ones selected can be seen in Fig.~\ref{fig:fig2}, where the Pearson and Kendall correlation heatmaps are plotted, for the numerical features.

\begin{figure}[htbp]
     \centering
     \subfloat[Pearson\label{fig:final_pearson}]{%
         \includegraphics[width=0.23\textwidth]{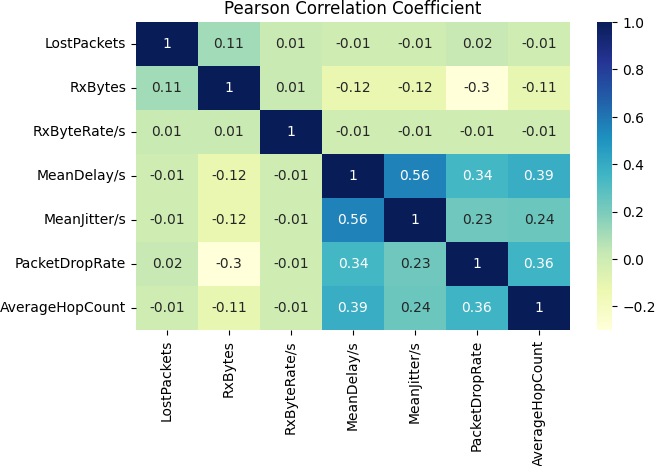}%
     }%
     \hspace{2pt} 
     \subfloat[Kendall\label{fig:final_kendall}]{%
         \includegraphics[width=0.23\textwidth]{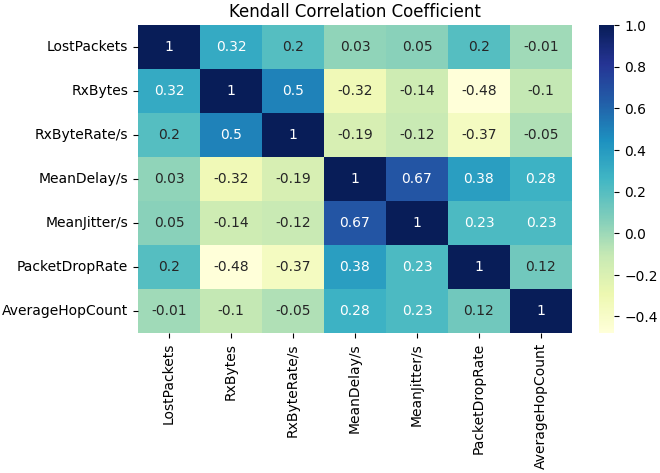}%
     }%
     \caption{Correlation Coefficients after Dropping Correlated Features for SHAP}
     \label{fig:fig3}
\end{figure}

From these we observe some high correlations, that need to be removed for the SHAP analysis, to avoid misleading results. An example is the use of the variables "TxByteRate/s", "TxPacketRate/s", "RxByteRate/s" and "RxPacketRate/s", in \cite{b12}, which share a non-linear correlation $>0.7$, that raises the danger of multi-colinearity, for false interpretations. We consider a threshold of $>=0.7$ and remove the ones that exceed it, thus leading us to the below correlation heatmaps in Fig.~\ref{fig:fig3}.
     
The remaining features can be seen below in Tab.~\ref{table:1}
\begin{table}[htbp]
  \centering
  \caption{Explainable Intrusion Detection Selected Features}
  \label{table:1}
  \begin{tabular}{|c|}
    \hline
    \textbf{\textit{Selected Features}}  \\
    \hline
    LostPackets, RxBytes, RxByteRate/s, \\ MeanDelay/s, MeanJitter/s, PacketDropRate, \\ AverageHopCount, DstPort654  \\
    \hline
  \end{tabular}
\end{table}

\subsection{Intrusion Detection}
For the classification of the tabular data, our experiments are concentrated in tree-ensembles, like bagging and boosting, deep neural networks, stacking ensembles of deep neural networks and hybrid stacking ensemble models that combine trees, ensembles and linear models, with Gaussian Naive Bayes. The baseline models are Logistic Regression and an SVM. The tree ensembles are XGBoost, XGBoost with Random Forest, Random Forest, LightGBM, Extra-Trees, Gradient Boosting, Histogram-Based Gradient Boosting, Decision-Tree-Based AdaBoost, Decision-Tree-Based Bagging. The first hybrid stacking model (Stacking-1) that we chose contains a Decision Tree, a Histogram-Based Gradient Boosting and a Gaussian Naive Bayes as base estimators and Ridge as the final estimator. The second hybrid model (Stacking-2) contains a Decision Tree, a Logistic Regression and a Gaussian Naive Bayes as base estimators and Ridge as the final estimator. The neural networks are MPL and RealMLP. Also the following stacking ensemble versions of deep neural networks RealMLP, NNFastAiTabular and NN-Torch, from the AutoGluon and PyTabKit libraries have been selected \cite{b4}, \cite{b15}. Grid-search, along with stratified $10$-fold cross validation was combined for hyperparameter tuning. After selecting the optimal hyperparameters, each of the pre-mentioned models was trained with stratified $10$-fold cross validation, with probability calibration, while holding a $4\%$ value range threshold for considering a model stable per evaluation metric.
\par The evaluation metrics, with which we evaluated the performance of each algorithm are F1, Precision, Recall, and Roc-Auc. From the metric results, in Tab.~\ref{table:2} and Tab.~\ref{table:3}, we observe that all tree-ensembles achieve score over $92\%$, which explains their superiority, with XGBoost being the best-performing of it's type. Furthermore, we observe very good to excellent performances, from the neural networks and to be more specific, of RealMLP. The ensemble neural networks have good to outstanding results, with the Ensemble versions of RealMLP and NN-Torch being the ones that stand out. Regarding the two baseline models, Logistic Regression struggles, while the SVM has a nearly-acceptable performance. 

\begin{table}[htbp]
  \centering
  \caption{Train/Test Precision and Recall Results ($\%$)}
  \label{table:2}
  \begin{tabular}{|c|c|c|c|c|}
    \hline
    \textbf{\textit{Model}} & \textbf{\textit{Train/Test-Precision}} & \textbf{\textit{Train/Test-Recall}} \\
    \hline
    XGBoost & 96.60/95.17 & 96.50/95.04 \\
    \hline
    XGBoost-RF & 95.58/94.56 & 95.49/94.44 \\
    \hline
    Random Forest & 95.75/94.17 & 95.68/94.09 \\
    \hline
    LightGBM & 95.19/94.27 & 94.95/94.01 \\
    \hline
    Extra-Trees & 92.90/91.20 & 92.52/90.80 \\
    \hline
    Gradient Boosting & 92.29/91.71 & 92.19/91.62 \\
    \hline
    HB-Gradient Boosting & 92.77/92.31 & 92.64/92.17 \\
    \hline
    AdaBoost(DT) & 95.91/94.85 & 95.18/93.93 \\
    \hline
    Bagging (DT) & 93.51/92.71 & 93.43/92.65 \\
    \hline
    Stacking-1 & 94.88/94.05 & 94.78/93.95 \\
    \hline
    Stacking-2 & 66.15/65.85 & 65.36/65.23 \\
    \hline
    MLP & 91.20/90.70 & 91.13/90.65 \\
    \hline
    RealMLP & 95.00/94.49 & 94.80/94.28 \\
    \hline
    Ensemble-RealMLP & 94.26/93.71 & 94.07/93.51 \\
    \hline
    Ensemble-NNFastAiTab & 86.63/86.30 & 85.69/85.34 \\
    \hline
    Ensemble-NN-Torch & 94.14/93.72 & 93.85/93.43 \\
    \hline
    SGD-SVM & 74.46/74.21 & 72.73/72.40 \\
    \hline
    Logistic Regression & 49.20/49.50 & 48.74/48.90 \\
    \hline
  \end{tabular}
\end{table}
\begin{table}[htbp]
  \centering
  \caption{Train/Test F1 and Roc-Auc Results  ($\%$)}
  \label{table:3}
  \begin{tabular}{|c|c|c|c|c|}
    \hline
    \textbf{\textit{Model}} & \textbf{\textit{Train/Test-F1}} & \textbf{\textit{Train/Test-Roc-Auc}} \\
    \hline
    XGBoost & 96.50/95.04 & 97.77/96.85 \\
    \hline
    XGBoost-RF & 95.49/94.45 & 97.14/96.48 \\
    \hline
    Random Forest & 95.68/94.09 & 97.26/96.26 \\
    \hline
    LightGBM & 94.95/94.02 & 96.80/96.21 \\
    \hline
    Extra-Trees & 92.56/90.84 & 95.27/94.18 \\
    \hline
    Gradient Boosting & 92.19/91.62 & 95.08/94.72 \\
    \hline
    HB-Gradient Boosting & 92.65/92.19 & 95.36/95.07 \\
    \hline
    AdaBoost(DT) & 95.15/93.92 & 96.93/96.15 \\
    \hline
    Bagging (DT) & 93.44/92.65 & 95.86/95.37 \\
    \hline
    Stacking-1 & 94.79/93.96 & 94.78/93.95 \\
    \hline
    Stacking-2 & 62.77/62.68 & 78.18/78.10 \\
    \hline
    MLP & 91.15/90.66 & 94.40/94.09 \\
    \hline
    RealMLP & 94.80/94.28 & 96.71/96.38 \\
    \hline
    Ensemble-RealMLP & 94.06/93.51 & 96.24/95.89 \\
    \hline
    Ensemble-NNFastAiTab & 85.90/85.56 & 90.98/90.76 \\
    \hline
    Ensemble-NN-Torch & 93.87/93.44 & 96.12/95.86 \\
    \hline
    SGD-SVM & 73.19/72.87 & 82.82/82.61 \\
    \hline
    Logistic Regression & 46.93/47.13 & 67.80/67.89 \\
    \hline
  \end{tabular}
\end{table}

\begin{figure}[htbp]
\centerline{\includegraphics[width=0.80\columnwidth]{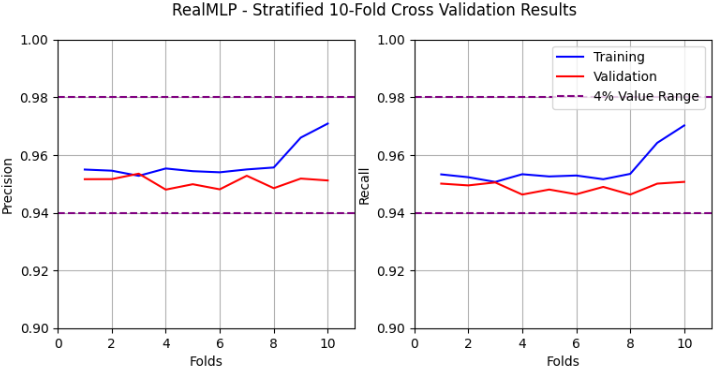}}
\caption{RealMLP Stratified 10-Fold CV.}
\label{fig4}
\end{figure}

To highlight the value of reliable training through stratified cross-validation we present the calibrated cross-validation results, of RealMLP, for precision and recall in Fig.~\ref{fig4}, where we observe stability, within the $4\%$ value range.

The confusion matrices for the training and testing set, for XGBoost, can be seen below in Fig.~\ref{fig5}. It is worth to mention that, both for in-sample and out-of-sample data, only a few Wormhole attacks trick the XGBoost, to misclassify them as Normal traffic. We can also see how XGBoost confuses some Blackhole attacks for Wormhole attacks and vice versa. All of these findings will be presented and justified in the following section.

\begin{figure}[htbp]
\centerline{\includegraphics[width=0.85\columnwidth]{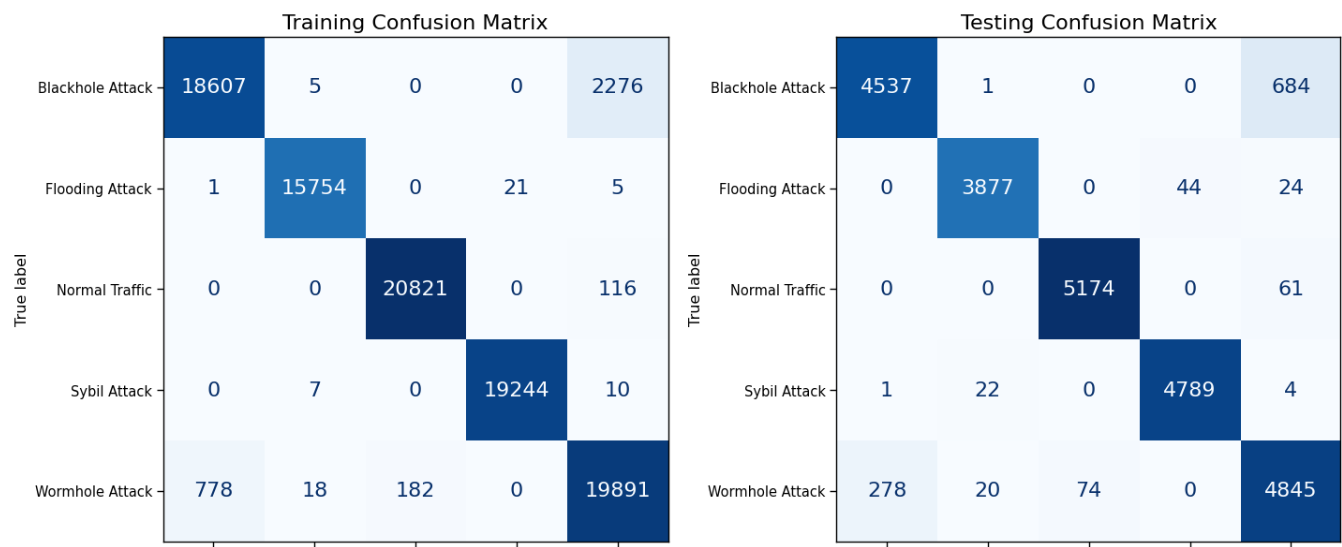}}
\caption{XGBoost Confusion Matrices}
\label{fig5}
\end{figure}

\section{Explainability}

To gain actual and trustworthy insight from our trained models, we will perform SHAP analysis, for our best model, XGBoost. Our Methodology starts by examining global and local explanations. Then we move to distribution analysis with violin plots, kernel density estimations and statistical divergence computation. Last but not least, we target the features that cause misclassifications, to get real insight about why and where our best model gets tricked. For most of our analysis, we utilized the SHAPASH framework.

\subsection{Global/Local Explanations}

\begin{figure}[htbp]
\centerline{\includegraphics[width=0.85\columnwidth]{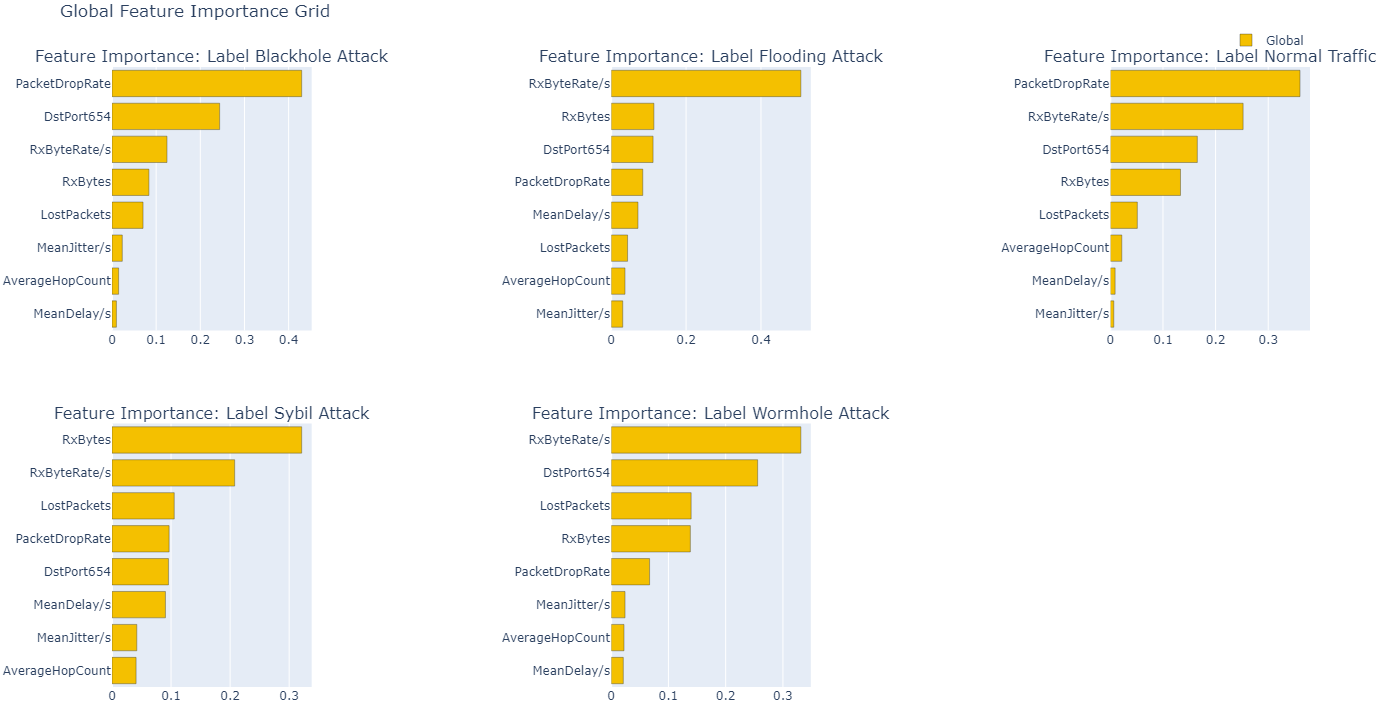}}
\caption{XGBoost Global Feature Importances per Label.}
\label{fig6}
\end{figure}

\begin{figure}[htbp]
\centerline{\includegraphics[width=0.85\columnwidth]{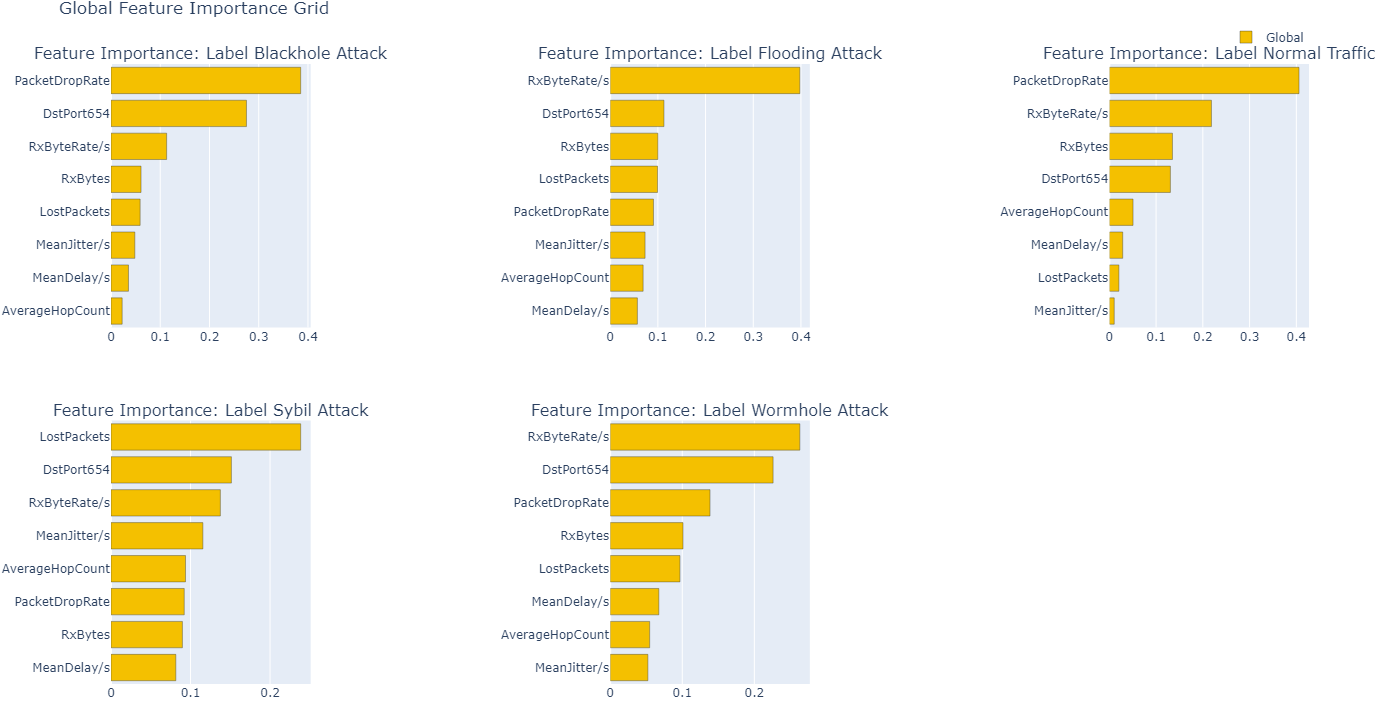}}
\caption{MLP Global Feature Importances per Label.}
\label{fig7}
\end{figure}

To understand the overall decision-making process of our model, XGBoost, we begin with global feature importances. In Fig.~\ref{fig6} we observe those, per label and it is clear that the features with the highest impact for most of the classes are "PacketDropRate", "RxByteRate/s", "RxBytes" and "DstPort654". From the same figure, we can understand, in-general, how each attack prioritizes the variables, that play a major role for a packet to be labeled as normal traffic (importance $>= 0.1$). Furthermore to compare XGBoost's efficiency with MLP, we present in Fig.~\ref{fig7} the global feature importances per label. From this we notice how the algorithm's decision are based obviously on the features mentioned above, but also at a higher percentage in the variables "MeanDelay/s", "AverageHopCount" and "MeanJitter/s", which causes the larger percentage of misclassifications.


\begin{figure}[htbp]
\centerline{\includegraphics[width=0.85\columnwidth]{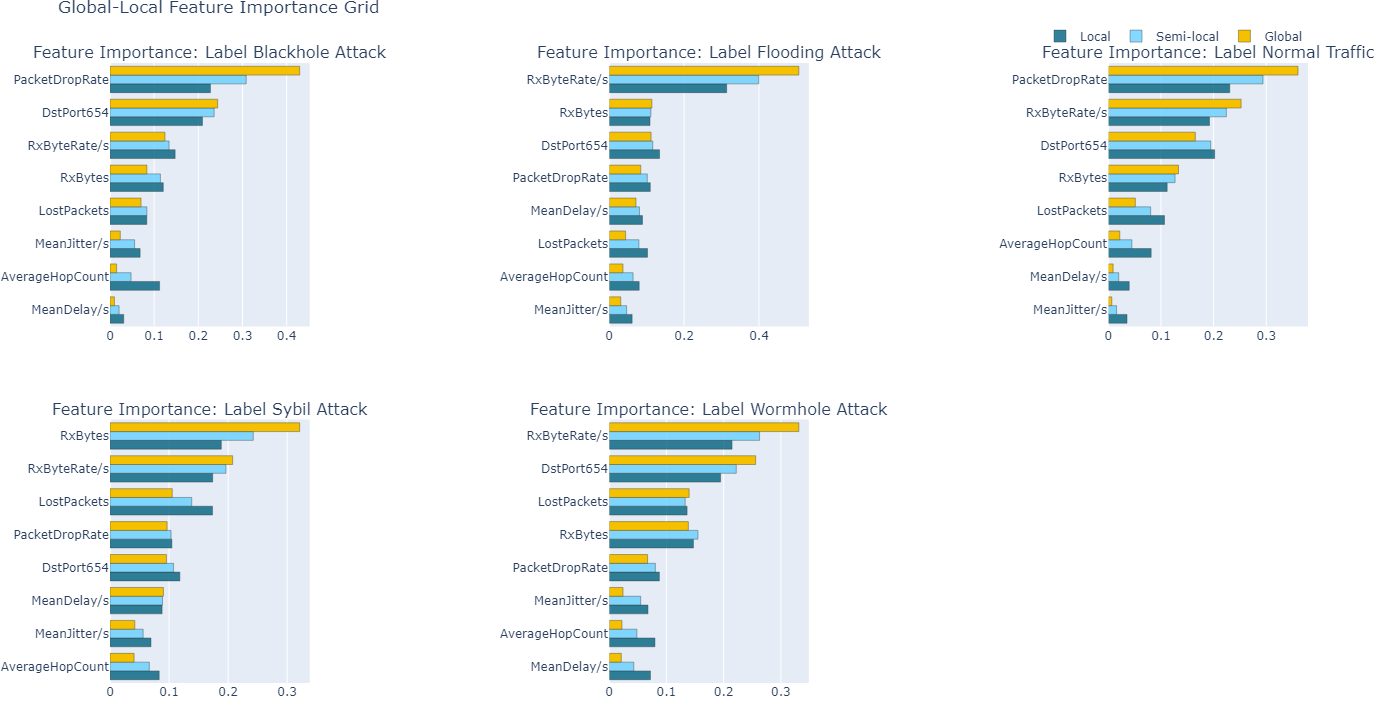}}
\caption{XGBoost Global, Local, Semi-Local Feature Importance per Label.}
\label{fig8}
\end{figure}

\begin{figure}[htbp]
\centerline{\includegraphics[width=0.85\columnwidth]{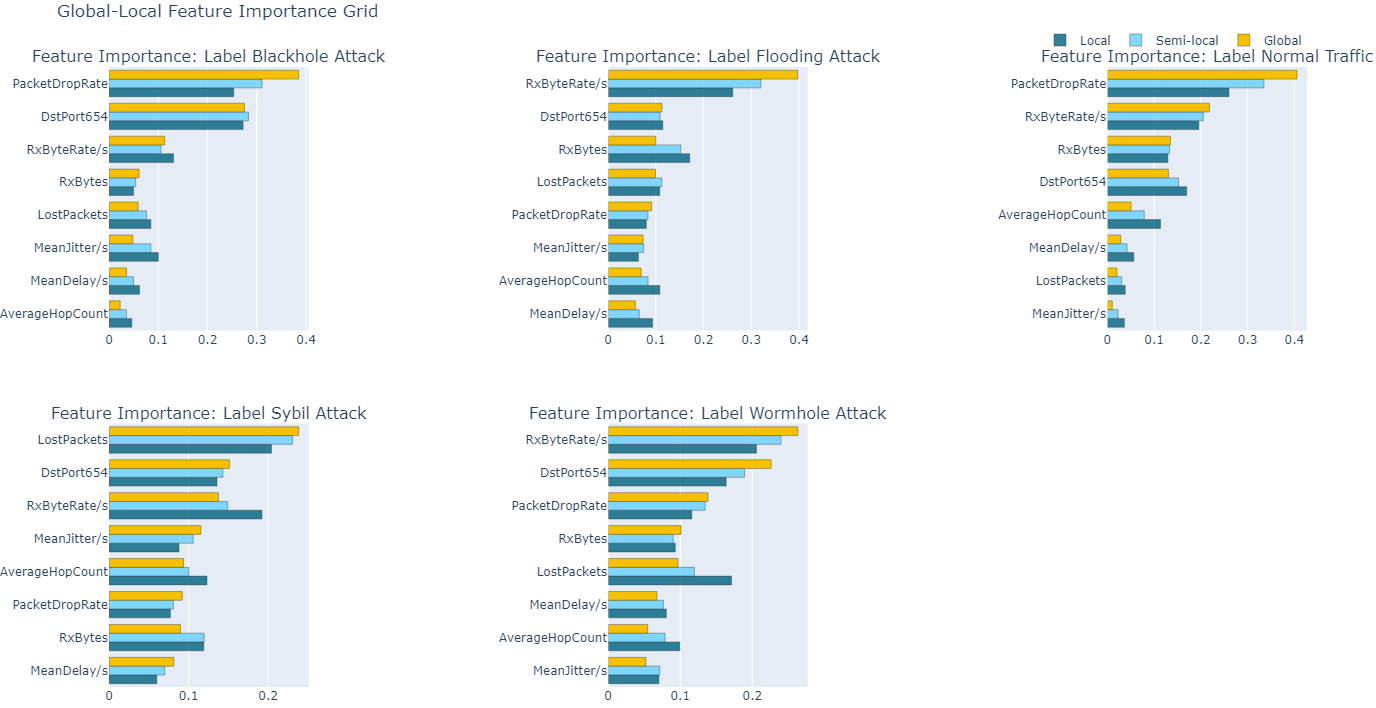}}
\caption{MLP Global, Local, Semi-Local Feature Importances per Label.}
\label{fig9}
\end{figure}

Moving to local explanations, as seen in Fig.~\ref{fig8}, for our best model, we see, if we look closely, how the importances in the variables "MeanDelay/s", "AverageHopCount" and "MeanJitter/s" have raised compared to the global ones. These findings prove that, misclassifications may occur, due to the fact that certain features, per label always, have larger percentage of contribution to some individual observations. As we can see from Fig.~\ref{fig9}, for MLP, the local feature importances to these three pre-mentioned variables, is even larger and more specifically in Sybil and Wormhole attacks. That justifies the lower performance of MLP, compared to XGBoost.


\subsection{Shape Explanations}

Below in Fig.~\ref{fig10} the Box-Plots per label can be seen and we notice the large number of outliers that exist in some variables. It is significant to mention, how similar is the dispersion of the outliers, for the features "MeanDelay/s", "AverageHopCount" and "MeanJitter/s" in the Wormhole and Blackhole attacks, the ones, in which the misclassifications occur in XGBoost, between these two classes. Their similarities, in terms of density, skewness and center of mass, is highlighted, with the aid of violin plots in Fig.~\ref{fig11}. This reveals that, our best model has room for improvement, in similar heavy outlier tails, between globally less important features. Such misclassification scenarios, for the dilemma of Blackhole or Wormhole, are avoided when the variable "PacketDropRate" obtains high values. Otherwise with values close to the median the model becomes hesitant. False predictions, between these two classes, when the model exists in ambiguity, due to the pre-mentioned reasons, can obviously arise, from the outliers of "DstPort654", as seen in the Wormhole attack.

\begin{figure}[htbp]
\centerline{\includegraphics[width=0.95\columnwidth]{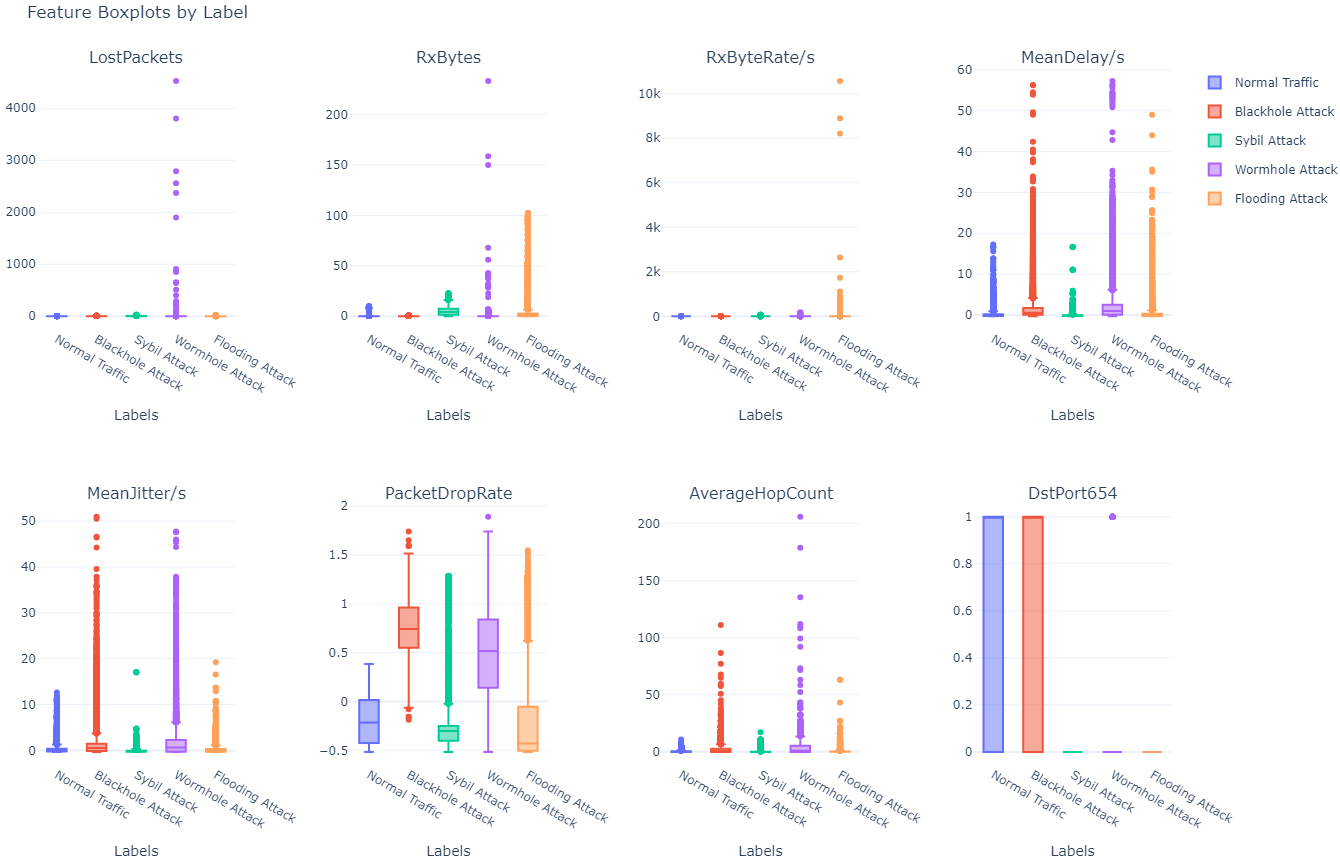}}
\caption{Feature Box-Plots per Label.}
\label{fig10}
\end{figure}

An easy way to analyze the shapes of our variables, along with their skewness, are the violin plots as seen in Fig.~\ref{fig11}. For each violin plot, the positive values push the probability of prediction, higher, towards the label, to which the plot corresponds to. The negative values lower the probability. In some features we see extremely long right tails in most of the classes and in combination with the center of mass being concentrated near zero, in the y-axis, we  can understand that extreme events, have very high impact on the model's decision. For example in the Flooding attack, all variables, but "RxByteRate/s", have long tails in their violin plots and wide center of mass around zero, meaning that rare observations make the difference in XGBoost's decisions. For the same attack, it is also interesting, that in the variable "RxByteRate/s", we have observations that aid both in the increase and decrease of the probability, but then it's heavy right tail slightly thickens at the end, which indicates a second peak in it's probability density function. The same we observe in the Wormhole attack for variable "RxByteRate/s". The data-points located in that spot, rule the model's decisions, towards labeling a packet as Flooding. Furthermore by comparing the violins, in features "MeanDelay/s", "AverageHopCount" and "MeanJitter/s", between Blackhole and Wormhole attacks, we notice their extreme shape similarity, in terms of density, thickness and long right tails, which supports the previous findings on the global and local explanations. This tells us that these features probably confuse the model, for these classes, because of their high similarity. About the two peaks mentioned above, we detect this phenomenon in other features and more specifically in labels Flooding, Sybil, Blackhole attack, but also in Normal traffic. These are indications of multi-modal distributions. To conclude with the violin plots, our best model is capable of identifying a large amount of rare events and make clever decisions, without being tricked by the high variance of our data and their multi-modal densities, in most cases. 

\begin{figure}[htbp]
\centerline{\includegraphics[width=0.90\columnwidth]{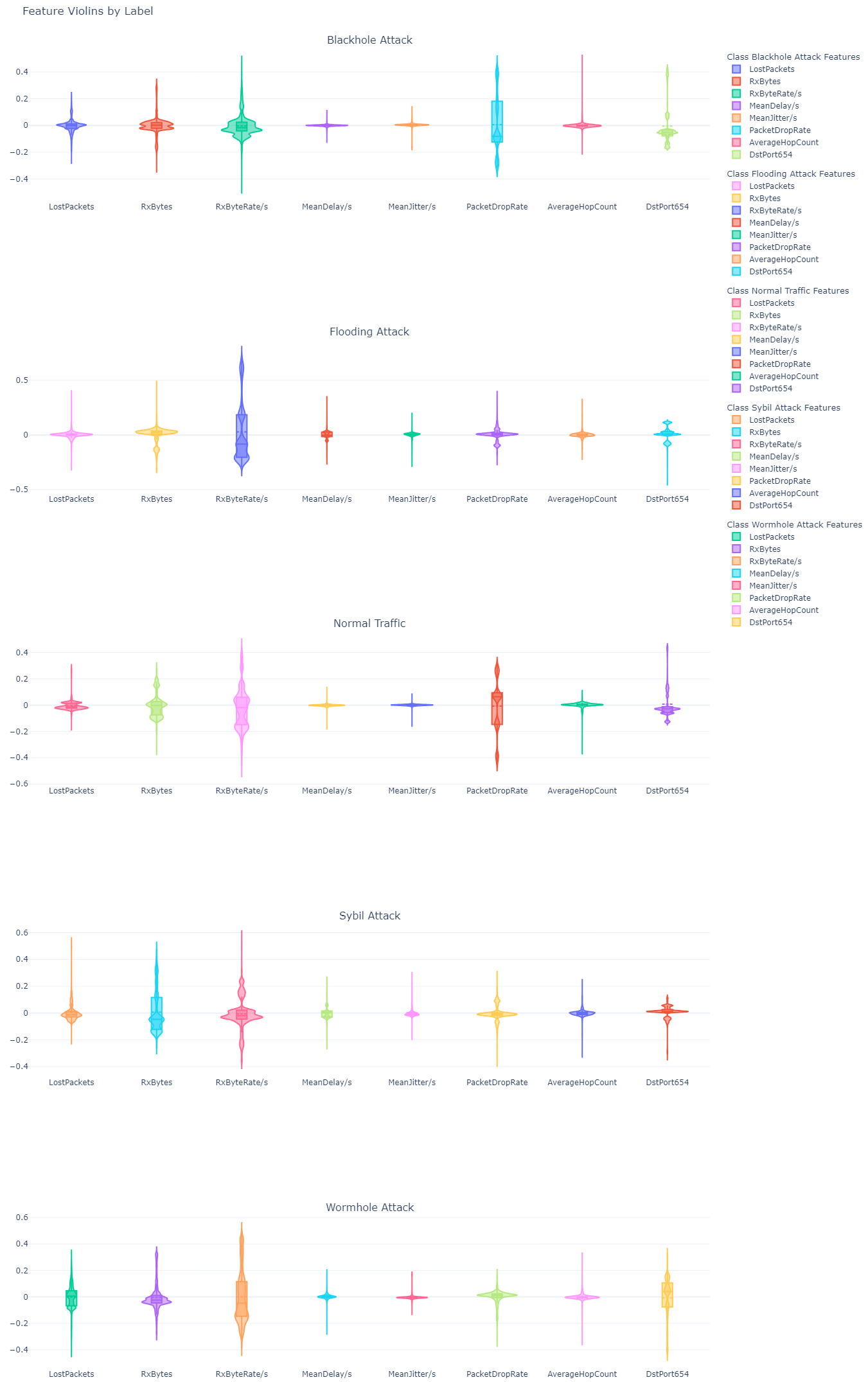}}
\caption{XGBoost Feature Violin Plots per Label.}
\label{fig11}
\end{figure}

\begin{figure}[htbp]
\centerline{\includegraphics[width=0.80\columnwidth]{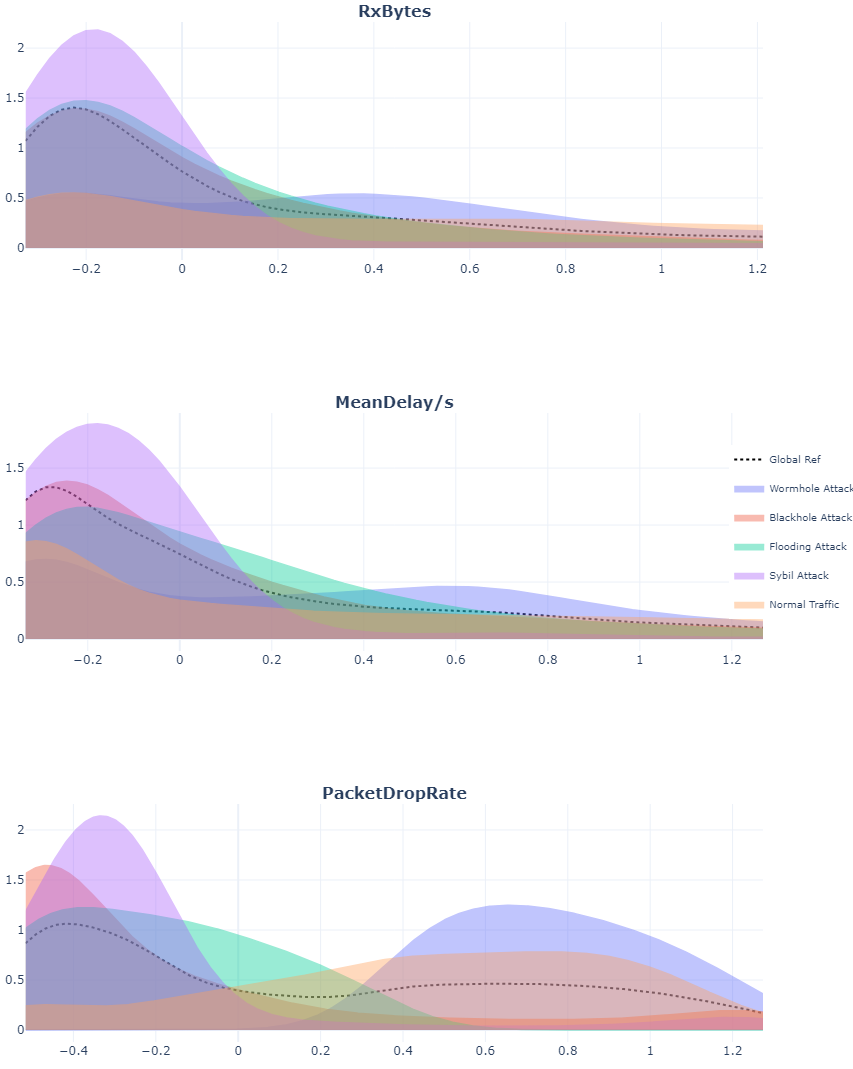}}
\caption{KDEs per Label}
\label{fig12}
\end{figure}

\begin{figure}[htbp]
\centerline{\includegraphics[width=0.80\columnwidth]{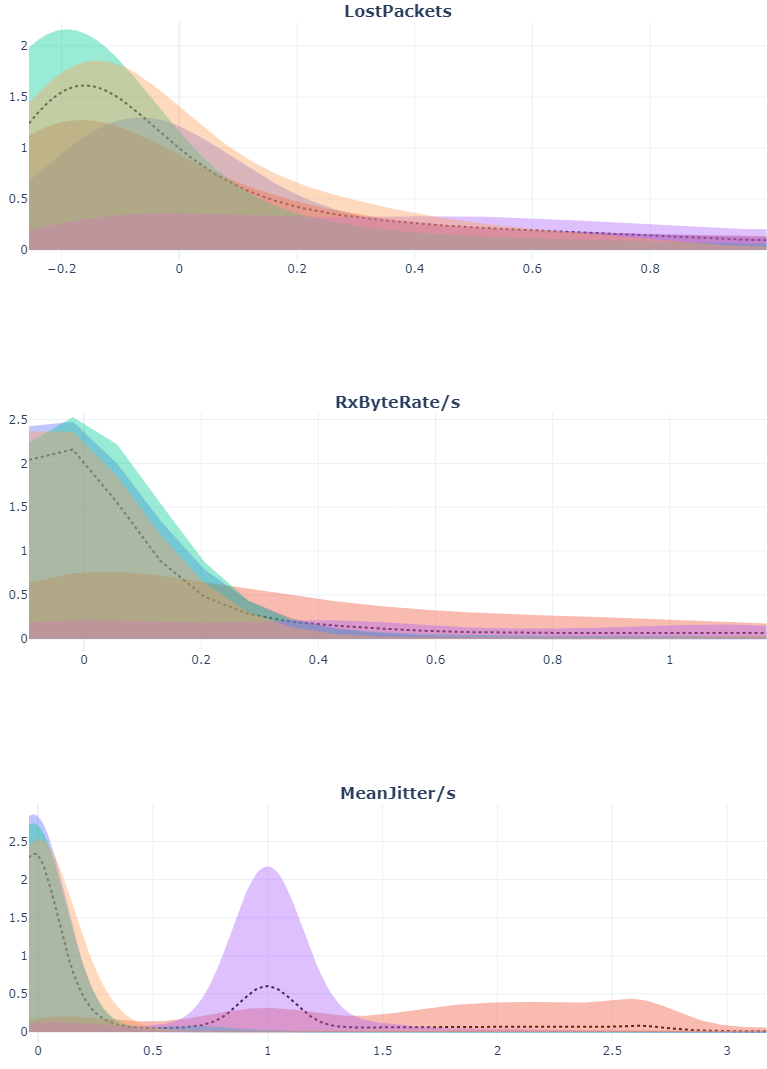}}
\caption{KDEs per Label}
\label{fig13}
\end{figure}

We applied Kernel Density Estimation, with Gaussian kernel and Scott bandwidth, as a visualization method to get a first general idea \cite{b25}. The visualizations can be seen in Fig.~\ref{fig12} and Fig.~\ref{fig13}. From those, we clarify long and heavy tails, as well as bimodal distributions. Additionally, we highlight, how similar some distributions are, in features like "LostPackets", "RxByteRate/s" and "RxBytes". 


\subsection{Uncertainty in Blackhole and Wormhole Attacks}

\begin{figure}[htbp]
\centerline{\includegraphics[width=0.70\columnwidth]{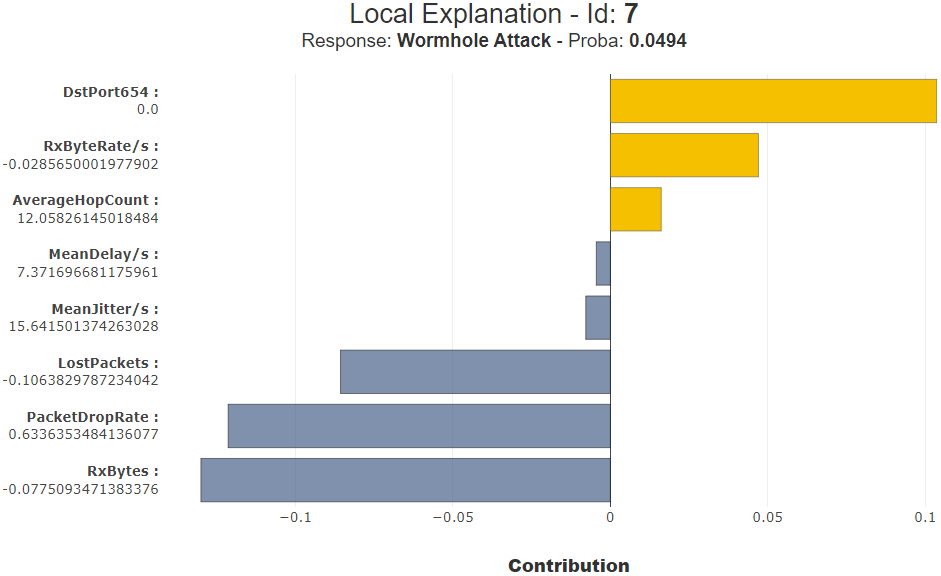}}
\caption{Local Explanation for the $7$th observation}
\label{fig14}
\end{figure}

To be more specific in our analysis, in Fig.~\ref{fig14}, we see the local explanation for the $7$th observation of our training dataset, which focuses on the Blackhole-Wormhole dilemma. This observation is a Wormhole attack, but is predicted as Blackhole. The variables "AverageHopCount", "MeanDelay/s" and "MeanJitter/s", all share high values, which is observed in both classes and it doesn't really help. The negative values observed in "LostPackets", stand for values below and close to the median of the distribution, after the transformation with the Robust scaler, which are indications of Blackhole attack. This because from both the Box-Plots and Violin plots in Fig.~\ref{fig10} and Fig.~\ref{fig11}, respectively, we see longer tails in this variable, for Wormhole attacks. The only clear indications for it, to be labeled as Wormhole is the destination port with value $0$. From the differentiator variable, "PacketDropRates", we see a value of $0.63$, which means it is located ahead of the median, a clue that supports the Blackhole scenario, as seen by the right heavy tail in the Box-Plot in Fig.~\ref{fig10}. 

\begin{figure}[htbp]
\centerline{\includegraphics[width=0.70\columnwidth]{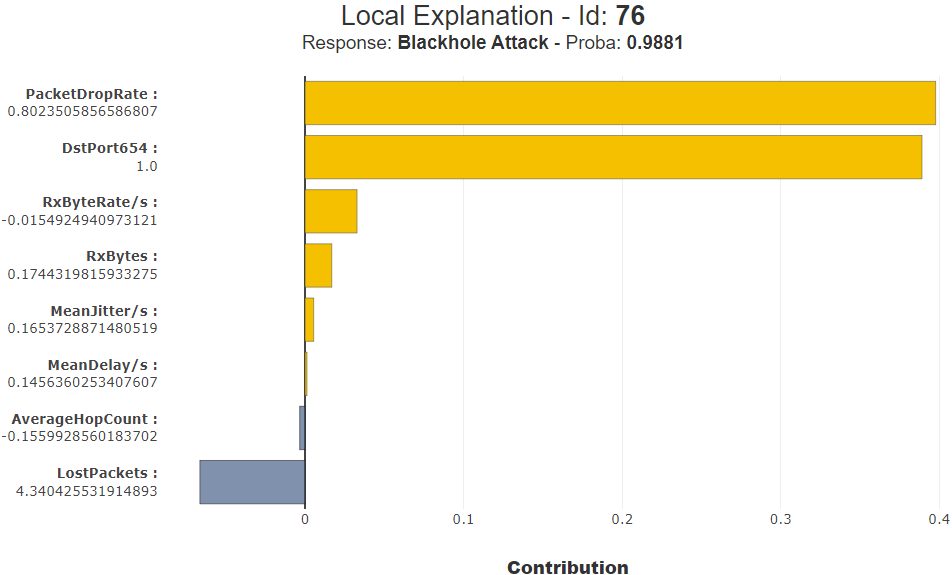}}
\caption{Local Explanation for the $76$th observation}
\label{fig15}
\end{figure}

In Fig.~\ref{fig15}, we highlight, how a solid decision looks like for a Blackhole attack. First the variable "PacketDropRate", has a high value, ahead of the median, leaning in favor of Blackhole. From the value $1$ in "DstPort654" we observe the same, as well as from the "RxBytes", because it is located close to the median. The only suspicion that could probably rise is the value of "LostPackets", because such high values are observed in the Wormhole Attacks, as it can be verified from Fig.~\ref{fig10}.

\subsection{Density Estimation and Westfall-Young Permutation Test}

To gain deeper statistical insights about our model's decision's we will perform the Westfall-Young Permutation Test, for a total of $1000$ permutations \cite{b16}, \cite{b17} and Kernel Density Estimation with Bandwidth optimization \cite{b25}. Due to the misclassifications that occurred, in the confusion matrix, between the classes Blackhole and Wormhole, we must search for the cause. Let $P_{i,k}$ be the distribution of the $k$-th class, for the $i$-th feature. For each of our continuous features, $i = 1, 2,..., p$, with $p=7$ and our two selected classes, $k = V, W$, we can formulate the hypothesis seen in ``$(1)$'', which addresses the equality of two density estimations. In our Max-Statistic test we have the following hypothesis, as seen in ``$(2)$'', hence it is called multiple testing \cite{b16}, \cite{b17}, for a total of $i$ features and two classes $W$ and $V$:

\begin{equation}
H_0: P_{i,V} = P_{i,W} \hhh against \hhh H_1: P_{i,V} \neq P_{i,W}
\end{equation}
\begin{equation}
\begin{gathered}
H_0: P_{i,V} = P_{i,W}, \hh i\in\{1,2,\ldots,7\}  \hhh against \hhh \\
H_1: P_{i,V} \neq P_{i,W}, \hh \exists i\in\{1,2,\ldots,7\}
\end{gathered}
\end{equation}

Because we will address only two categories, consider $P_{j,W} = Q_{i}$. Both $P_{i}(x), Q_{i}(x)$ are estimated with Kernel Density Estimation, with Gaussian kernel and an embedded cross validation, for optimal bandwidth selection \cite{b25}, notated as $\hat{P}_{i}(x), \hat{Q}_{i}(x)$. For this procedure, we select as a test statistic the Jensen-Shannon distance (Square Root of Jensen-Shannon divergence):

\begin{equation}
\begin{gathered}
T_{i} = JS_{Dist}\Big(\hat{P}_{i}, \hat{Q}_{i}\Big) = \sqrt{JS_{Div}\Big(\hat{P}_{i}, \hat{Q}_{i}\Big)} = \\
= \sqrt{\frac{1}{2}D_{KL}\Big(\hat{P}_{i}, \hat{M}_{i}\Big) + \frac{1}{2}D_{KL}\Big(\hat{Q}_{i}, \hat{M}_{i}\Big)}, \hh with \\
D_{KL}\Big(\hat{P}_{i}, \hat{M}_{i}\Big) = \sum_{x\in \{x_B,x_W\}}\hat{P}_{i}(x)\log_2\Bigg(\frac{\hat{P}_{i}(x)}{\hat{M}_{i}(x)}\Bigg), \\
\hat{M}_{i}(x) = \frac{1}{2}\Big(\hat{P}_{i}(x) + \hat{Q}_{i}(x)\Big), \hhh i=1,2,\ldots,p
\end{gathered}
\end{equation}

The Max-Statistic for the $b$-th permutation is:

\begin{equation}
\begin{gathered}
T_{b}^{JSD} = \max\{T_{1b}, T_{2b}, \ldots, T_{pb}\}, \\ with \hh b = 1, 2, \ldots, B, \hh and \hh B=1000
\end{gathered}
\end{equation}

The FWER-adjusted $p-value$ of the test, for each feature can be calculated as seen in ``$(5)$'':
\begin{equation}
\begin{gathered}
p-value_{i} = \frac{\sum_{b = 1}^{B}I_{T_b^{JSD}}+1}{{B + 1}}, \hhh with \hhh \\
I_{T_b^{JSD}} = 1 \hh if \hhh T_b^{JSD}\geq T_{i} \hh else \hh 0
\end{gathered}
\end{equation}

For our experiments we consider a significance level of $\alpha = 5\%$ and focus on the Blackhole-Wormhole dilemma. Since $B=1000$, $p=7$, $k=2$, we have the permutation test results in Tab.~\ref{table:4}. 

\begin{table}[htbp]
  \centering
  \caption{Permutation Test Results}
  \label{table:4}
  \begin{tabular}{|c|c|c|c|c|}
    \hline
    \textbf{\textit{Feature}} & \textbf{\textit{Jensen-Shannon Distance}} & \textbf{\textit{$p-value$}} \\
    \hline
    LostPackets & 0.376153 & $<$0.001 \\
    \hline
    RxBytes & 0.743254 & $<$0.001 \\
    \hline
    RxByteRate/s & 0.197168 & 1.000000 \\
    \hline
    MeanDelay/s & 0.546946 & $<$0.001 \\
    \hline
    MeanJitter/s & 0.546824 & $<$0.001 \\
    \hline
    PacketDropRate & 0.825746 & $<$0.001 \\
    \hline
    AverageHopCount & 0.539645 & $<$0.001 \\
    \hline
  \end{tabular}
\end{table}

\begin{figure}[htbp]
\centerline{\includegraphics[width=0.85\columnwidth]{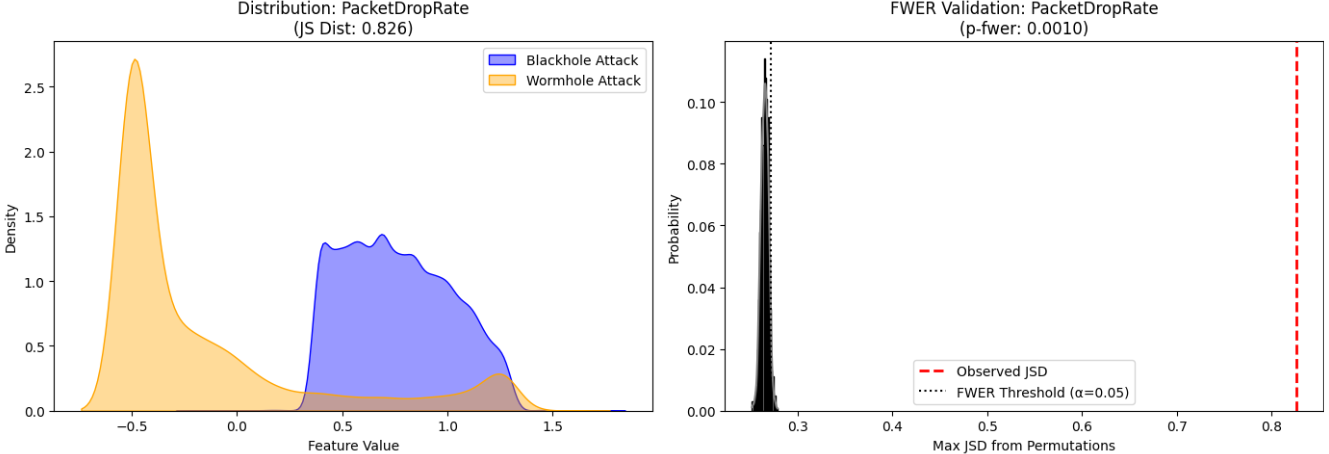}}
\caption{Feature: PacketDropRate}
\label{fig16}
\end{figure}

From Tab.~\ref{table:4} we have $p-valaue < a$ for all features, except "RxByteRate/s", which leads us to reject $H_0$ for those features, meaning that their estimated KDEs are different at a significance level of $a = 5\%$, but does not guarantee lack of overlap, of individual points. In the figures below, we present a few of the estimated densities, for each class, on the left and the $p-values$ of the test, on the right. In order to analyze the misclassifications we must look at Fig.~\ref{fig16}, where we can see clearly, the two densities of feature "PacketDropRate". We notice the bimodal-density for the Wormhole category, where the smaller peak, seems to mimic the behavior of Blackhole attacks, in the overlap zone $[0.25,1.4]$. High overlap percentage, in certain areas, share  also the variables "MeanJitter/s", "MeanDelay/s" and "AverageHopCount". The phenomenon of Density Support Intersection, the overlap in the plots, is the one that confuses XGBoost, in so many features, due to absence of linear separability, even when the estimated densities seem statistically different, in terms of shape, with the Max-Statistic of Jensen-Shannon. The problem with this dataset is that even the best feature for distinguishing Blackhole from Wormhole attacks has a percentage of density overlap.

\begin{figure}[htbp]
\centerline{\includegraphics[width=0.85\columnwidth]{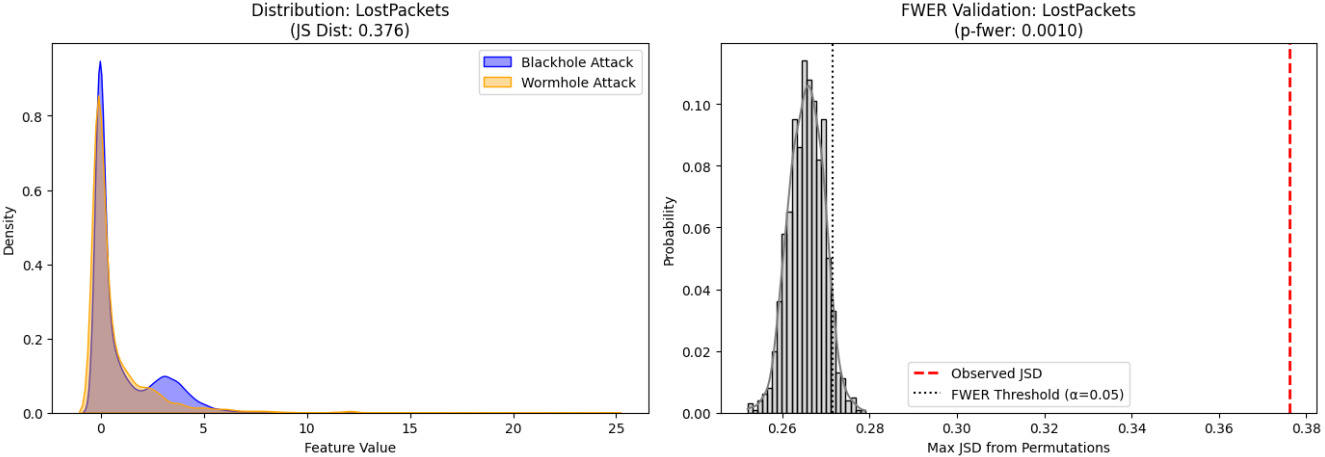}}
\caption{Feature: LostPackets}
\label{fig17}
\end{figure}

\begin{figure}[htbp]
\centerline{\includegraphics[width=0.85\columnwidth]{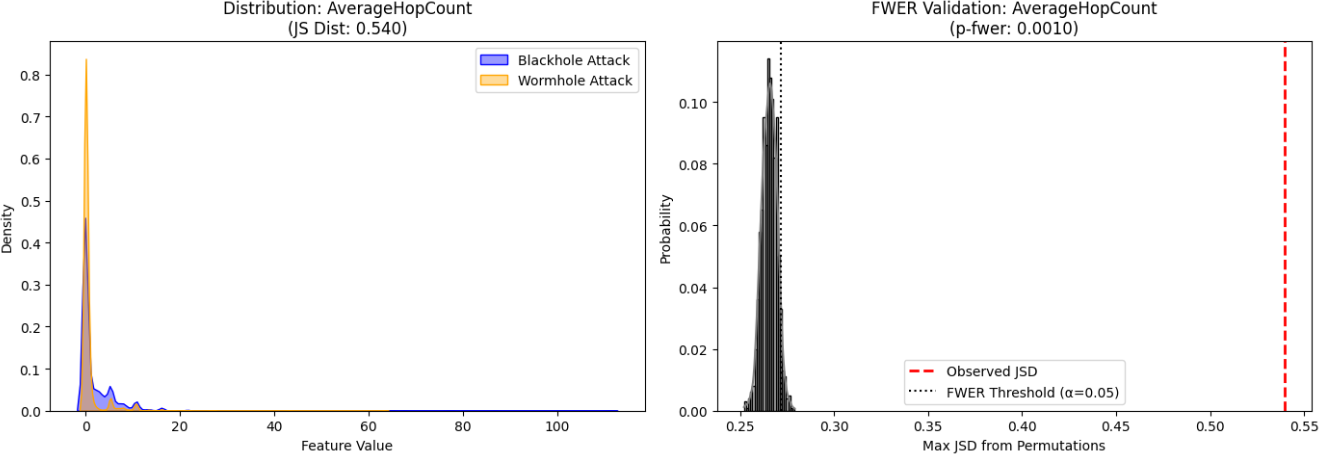}}
\caption{Feature: AverageHopCount}
\label{fig18}
\end{figure}

\begin{figure}[htbp]
\centerline{\includegraphics[width=0.85\columnwidth]{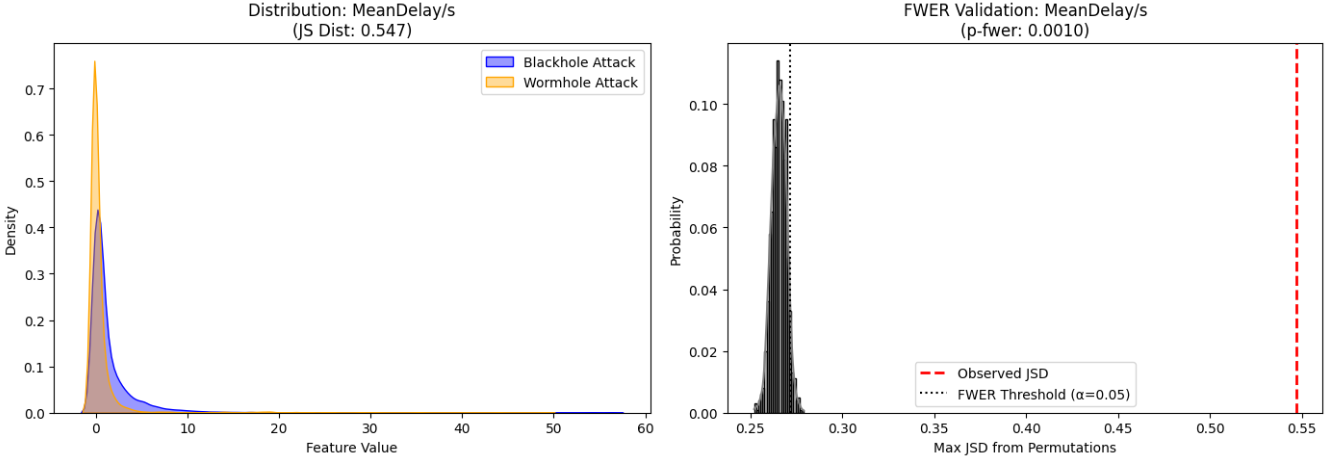}}
\caption{Feature: MeanDelay/s}
\label{fig19}
\end{figure}

\begin{figure}[htbp]
\centerline{\includegraphics[width=0.85\columnwidth]{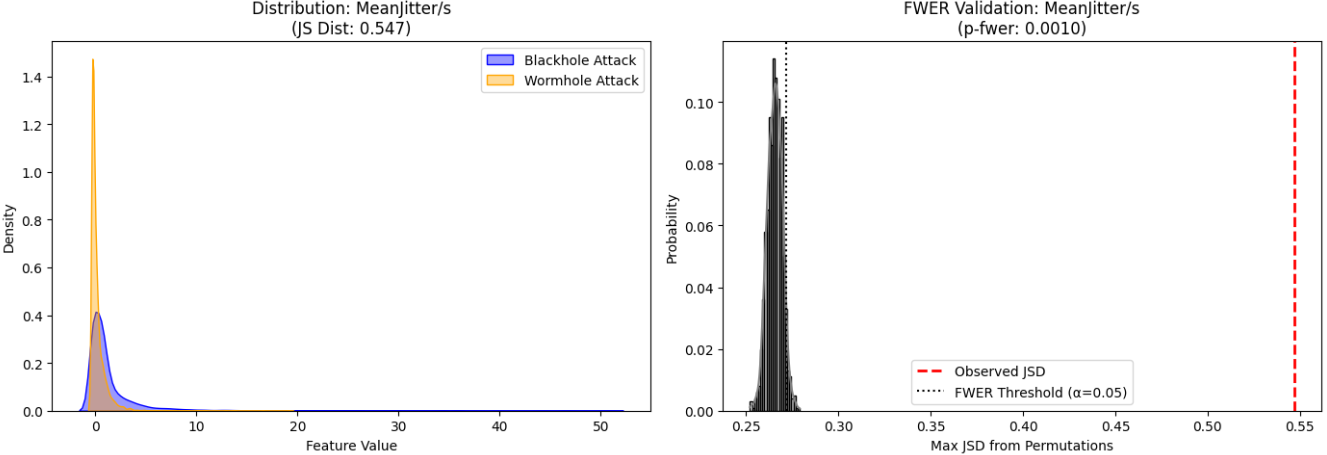}}
\caption{Feature: MeanJitter/s}
\label{fig20}
\end{figure}

\section{Conclusions}

In this paper we applied best practices for data pre-processing and provided top-performing algorithms and hybrid models, that are capable to detect intrusions in UAV systems, compared to baseline models. After the selection of our best model, we move to SHAP analysis, where we establish solid proof of our model's internal decision mechanisms, with local and global explanations, that help us clarify, which variables contribute to the predictions of the general tendency and which to individual data-points, per class. We didn't simply describe the general idea, but gave specific examples of true and false predictions. With the visual density and shape analysis we analyzed and explained our data's skewed, multi-modal densities and the false predictions that occur, due to the Support Intersection between Blackhole and Wormhole attacks. With the combination of the Westfall-Young test, the KDE and the JSD, we distinguish the different nature of the attacks, even with the high percentage of density overlap. These are proof that XGBoost's false predictions, do not happen, because of weak training or data pre-processing, but because of the massive support intersection, in UAVIDS-2025. Even with this major problem our best model has an outstanding performance. By combining all the above techniques, we have a solid view of the attacks. Future work includes cross-dataset validation, with other well-known UAVIDS datasets, that share the same feature space, use of Deep-CNNs with $2$D Convolutions for intrusion detection to capture the spatial characteristics and RNNs to handle the dataset as online training. Last but not least application of Density Overlap solutions. 

\section{Acknowledgments}
This work has been partially funded by the ‘advaNced cybErsecurity awaReness ecOsystem for SMEs’ (Nero) project, which has received funding from the European Union’s Digital Europe Programme (DEP) under grant agreement No. 101127411. Prof. Christos Douligeris would like to thank Prof. E. Iakovou and the Energy Institute of the Texas A$\&$M University for their hospitality.

\end{document}